\documentstyle[aps,prl,epsf,twocolumn,epsfig,floats]{revtex}

\begin{document}
\draft


\wideabs{

\title{An angle-resolved photoemission spectral function analysis of the electron doped
cuprate Nd$_{1.85}$Ce$_{0.15}$CuO$_4$}

\author{N.P.  Armitage$^{1,2,3}$, D.H.  Lu$^{3}$, C.  Kim$^{3}$,
A.  Damascelli$^{3,4}$, K.M.  Shen$^{3,4}$, F.  Ronning$^{2,3}$,
D.L.  Feng$^{2,3}$, P.  Bogdanov$^{3,4}$, X.J.  Zhou$^{3,5}$, W.L.
Yang$^{5}$, Z.  Hussain$^{5}$, P.  K.  Mang$^{3,4}$, N.
Kaneko$^{3,4}$, M.  Greven$^{3,4}$, Y.  Onose$^{6}$, Y.
Taguchi$^{6}$, Y.  Tokura$^{6}$, and Z.-X.  Shen$^{2,3,4}$}

\address{$^{1}$Department of Physics and Astronomy, University of
California, Los Angeles, CA 90095}

\address{$^{2}$Department of Physics, Stanford University,
Stanford, CA 94305}

\address{$^{3}$Stanford Synchrotron Radiation Laboratory, Stanford
University, Stanford, CA 94305}

\address{$^{4}$Department of Applied Physics, Stanford University,
Stanford, CA 94305}

\address{$^{5}$Advanced Light Source, Lawrence Berkeley National
Laboratory, Berkeley, CA 94720}

\address{$^{6}$Department of Applied Physics, The University of
Tokyo, Tokyo 113-8656, Japan}

\date{\today}
\maketitle
\begin{abstract}

Using methods made possible by recent advances in photoemission
technology, we perform an indepth line-shape analysis of the
angle-resolved photoemission spectra of the electron doped
($n$-type) cuprate superconductor Nd$_{1.85}$Ce$_{0.15}$CuO$_4$.
Unlike for the $p$-type materials, we only observe weak mass
renormalizations near 50-70 meV.  This may be indicative of
smaller electron-phonon coupling or due to the masking effects of
other interactions that make the electron-phonon coupling harder
to detect.  This latter scenario may suggest limitations of the
spectral function analysis in extracting electronic self-energies
when some of the interactions are highly momentum dependent.

\end{abstract}
} 

\section{Introduction}

Angle-resolved photoemission spectroscopy (ARPES) is one of the
most direct probes of the electronic structure of solids.  In
recent years, with the advent of $Scienta$ analyzers and high-flux
beam lines, a new era of photoemission line-shape analysis has
been ushered in\cite{AndreaRMP,Argonne}.  These advances offer
unprecedentedly high momentum and energy resolution, as well as
the ability to perform simultaneous parallel angle scanning and
thereby generate direct two dimensional (2D) $E-{\vec k}$ images
of ARPES spectral functions. Whereas, ARPES data have been
displayed traditionally as energy distribution curves (EDC's) in
which the photoemission intensity is plotted as a function of
energy at specific angles, now they can be represented in terms of
these detailed 2D intensity plots and in analogy with EDC's,
complementary momentum distribution curves (MDC's) can be
generated.  Such advances are making new analysis methods
possible, where one is able to bring to bear the considerable
mathematical machinery of many-body physics for intuition and
interpretation.

Within the sudden approximation, ARPES measures the
single-particle spectral function $A({\vec
k},\omega)$\cite{AndreaRMP,Argonne,Hufner}.  Assuming that the
material under investigation has well-defined electronic
excitations \cite{FermiL}, the single-particle spectral function
can be represented compactly in terms of a complex self-energy
$\Sigma = \Sigma^{\prime}+i \Sigma^{\prime\prime}$ as

\begin{equation}
A({\vec k},\omega)=\frac{1}{\pi}\frac{\Sigma^{\prime\prime}({\vec
k},\omega)}{[\omega-\epsilon_{\vec k}-\Sigma^\prime({\vec
k},\omega)]^2+[\Sigma^{\prime\prime}({\vec k},\omega)]^2}
\nonumber
\end{equation}

In momentum regions where $\Sigma({\vec k},\omega)$ is a weak
function of ${\vec k}$ and the bare dispersion can be linearized
as $\epsilon_{\vec k}= {\vec v_{F}^{0} }\cdot ({\vec k}-{\vec
k_{F}})$ (where ${\vec v_{F}^{0} }$ is the bare band velocity)
$A({\vec k},\omega)$ at constant $\omega$ can be put in the form
of a Lorentzian centered at $\omega/ {\vec v_{F}^{0}} +
k_{F}-\Sigma^\prime({\vec k},\omega)/{\vec v_{F}^{0} }$ with width
$\Sigma^{\prime\prime}({\vec k},\omega)/{\vec v_{F}^{0} }$.

Fits to MDC's gives one, in principle, not only a measure of
$\Sigma^{\prime\prime}({\vec k},\omega)$, but also a more accurate
parameterizations of the renormalized dispersion in the case of a
rapidly changing $\Sigma^\prime({\vec k},\omega)$.  Such an
analysis allows one to probe the dominant kinds of
electron-electron interactions as well as for signatures of
electronic interaction with bosonic modes.  Such studies are an
area of much current focus in the ARPES community with, for
instance, the observation of bosonic effects in the ARPES spectra
of systems with strong electron-phonon coupling such as Be and Mo
\cite{Hengsberger,LaShell,VallaMo}.  Recently, similar features
have been found in the low-energy ARPES spectra of a number of
high-temperature hole doped cuprate superconductors.  A mass
renormalization, or ``kink'', in the dispersion has been found
ubiquitously in the $p$-type materials (at $\approx$ 70
meV)\cite{Pasha,Ale}; its origin is a matter of much current
debate.  Some groups have pointed to a many-body electronic
source\cite{Kaminski,Johnsonkink,Gromko}, that is perhaps related
to the magnetic resonance mode discovered via neutron scattering.
However, this assignment has been disputed on the grounds that the
resonant mode may have insufficient spectral weight to cause the
observed effect with reasonable estimates for its coupling
strength\cite{what_res_not_do,rebuttal}.  Others have argued that
the kink's presence above T$_{c}$ and its universality among
material classes demonstrate a phononic origin and indicates the
strong role that lattice effects have on the low-energy
physics\cite{Ale}.

\begin{figure*}[htb]
    \centering
    \includegraphics[width=18cm]{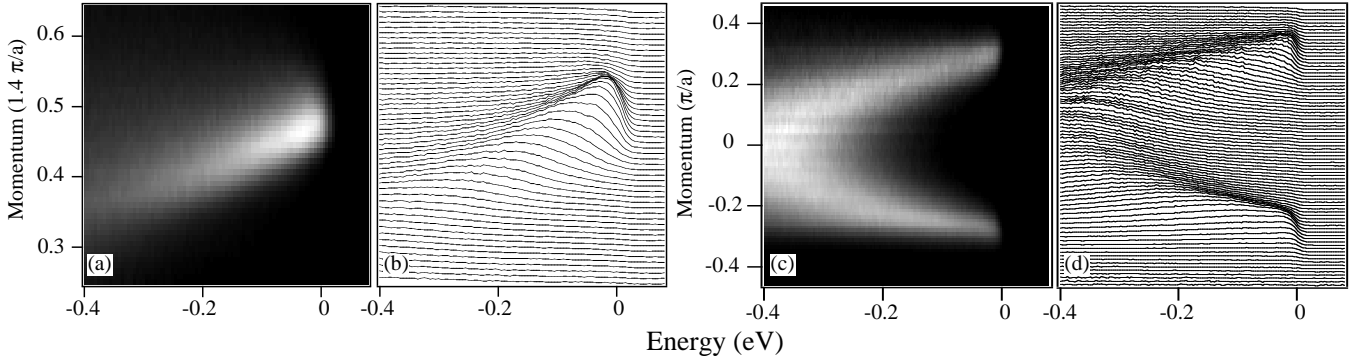}
    \vspace{.1cm}
    \caption{ Data from high symmetry directions with
    $E_{\gamma}$=55 eV (a) Image of the spectral function along
    the $\Gamma$ to $(\pi,\pi)$ direction.  (b) EDC's from (a).
    (c) Image of the spectral function along the $(\pi,-\pi)$ to
    $(\pi,\pi)$ direction.  (d) EDC's from (c).  The lower piece
    of the parabola in (c) is slightly distorted from its
    intrinsic shape as these data were at the edge of the detector
    multichannel plate and were likely affected by astigmatism in
    the electron lens.} \label{Fig1_Kink}
\end{figure*}

Although, in principle, it is straightforward with the above
analysis to extract the electronic self-energy from ARPES spectra,
it is not as obvious how to assign parts of the self-energy to
different origins.  For instance, is the phenomenon that causes
the kink the same one that causes the pseudogap in the underdoped
materials, or are multiple effects playing a role? Moreover, there
have even been reports that distinguish the kinks near the
$(\pi,0)$ and $(\pi/2,\pi/2)$ positions \cite{Gromko}.  Similar
manifestations of the various phononic, antiferromagnetic, charge
order, stripe, and structural effects (not to mention the
possibility of hidden orders \cite{DDW}) that may exist in the
cuprates, makes it difficult to extract unambiguous information
from a spectral function analysis.

Such problems may be particularly acute in the electron doped
system Nd$_{2-x}$Ce$_{x}$CuO$_4$\cite{Tokura}.  In previous ARPES
work on the ``optimally-doped'' ($x=0.15$) system, high-energy
pseudogap effects (i.e. the suppression of low-energy spectral
weight over an extended frequency range) similar to those in the
underdoped $p$-type compounds were observed\cite{estruct,NCCOGap}.
However, unlike in the $p$-type materials, these regions of
maximum pseudogap were not on the Fermi surface (FS) near
$(\pi,0)$ (the maximum of the $d_{x^{2}-y^{2} }$ functional form),
but instead at ``hot-spots'' at the intersection of the FS with
the antiferromagnetic Brillouin zone (BZ) boundary\cite{estruct}.
This, along with the fact that the antiferromagnetic and
superconducting phases in the highest-$T_{c}$ samples ($x=0.15$)
are in very close proximity to each other and may in fact
coexist,\cite{Yamadaneutron} led to the conclusion that this
high-energy pseudogap was due to the effects of antiferromagnetic
or other orders with characteristic wave vector
$(\pi,\pi)$\cite{estruct}. As the signatures of boson-electron
interactions are somewhat generic, it may be difficult to
distinguish interactions with different types of modes (for
instance phonons vs spin fluctuations) if the regions of the BZ,
where such effects are observed to be the strongest, overlap. In
the context of Eq. 1, this raises the question of the
$\vec{k}$-independence of the self-energy implied in this type of
analysis, especially in the $n$-type materials where the
momentum-dependent scattering appears to be very strong even near
optimal doping \cite{estruct}.

Keeping the above considerations in mind, we present an ARPES
line-shape analysis of the electron doped ($n$-type) cuprate
superconductor Nd$_{1.85}$Ce$_{0.15}$CuO$_4$.  We use essentially
same methodologies that have been employed in the study of the
$p$-type compounds as we believe that a direct comparison has
utility despite the fact that there may be possible complications.
We find that there is no kink along the zone diagonal, in contrast
to the $p$-type materials where this is the region of the BZ where
it can be most unambiguously identified.  In the $(\pi,0)$ to
$(\pi,\pi)$ direction, a very weak kink in the quasiparticle
dispersion can be found.  However, it is not straightforward to
identify this effect with phonons, since competing effects cannot
be separated unambiguously in this region of the BZ. Despite an
apparent lack of a mass renormalization along the zone diagonal, a
very slight drop in the scattering rate is observed in this
direction.  No such drop in the scattering rate is observed near
the $(\pi,0)$ to $(\pi,\pi)$ FS crossing.  As it is reasonable
that this class of many-body effects along the nodal direction in
the $p$-type compounds can be identified with effects of phonons,
the lack of a kink and weak drop in the scattering rate along the
same direction indicates a weaker electron-phonon interaction in
Nd$_{1.85}$Ce$_{0.15}$CuO$_4$.

\section{Experiment}

Angle-resolved photoemission data were recorded at Beam line
10.0.1.1 of the Advanced Light Source with a photon energy of 55
eV in a glancing-incidence geometry.  The polarization angles were
45$^{\circ}$ to the Cu-O bonds for the $\Gamma$ to $(\pi,\pi)$ cut
and along the Cu-O bonds for the $(\pi,0)$ to $(\pi,\pi)$ cut. The
energy resolution was typically 18 meV and angular resolution was
0.14$^{\circ}$ (corresponding to $\approx 0.5\%$ of the Brillouin
zone size).  This configuration is identical to that used in an
extensive study of the kink in the hole doped compounds\cite{Ale}.
The samples were cleaved $in$ $situ$ at low temperatures in vacuum
better than 4 x 10$^{-11}$ Torr.  As there are only minor changes
to the spectra when entering the superconducting
state\cite{NCCOGap}, all displayed spectra were taken at low
temperature ($\approx$ 15-25 K).  No signs of surface aging were
seen for the duration of the experiment ($\approx 24$ h).  In
addition to the displayed data, extensive measurements were
performed on Beam line 5-4 of the Stanford Synchrotron Radiation
Laboratory; results were obtained that are consistent with those
displayed here.

Single crystals of Nd$_{1.85}$Ce$_{0.15}$CuO$_4$ were grown by the
traveling-solvent grown method in 4 atm. of O$_{2}$.  The as-grown
crystals are not superconducting and must be annealed to remove
excess apical oxygen.  The displayed data come from samples that
were heated in flowing argon gas for 20 h  at 920$^{\circ}$C, and
then a further 20 h  at 500$^{\circ}$C under pure
O$_2$\cite{samples}. Resistivity and magnetic susceptibility
measurements show an onset of the superconducting transition at 24
K.

\section{Results}

\begin{figure}[htb]
    \centering
    \includegraphics[width=7.5cm]{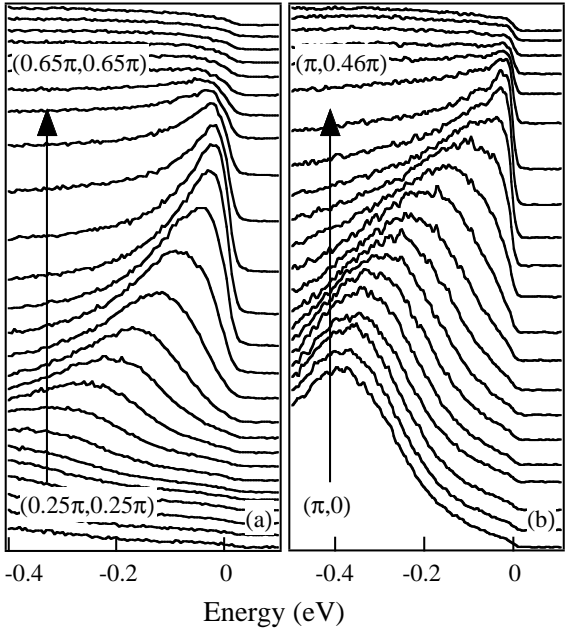}
    \vspace{.1cm}
\caption{$E_{\gamma}$=55 eV EDC's representing the data of Fig. 1
in a more traditional fashion.  (a) ($0.25\pi,
0.25\pi$)$\rightarrow$($0.65\pi, 0.65\pi$).  (b) ($\pi,
0$)$\rightarrow$($\pi, 0.46\pi$).  The EDC's are offset for
clarity. } \label{CloseEDC}
\end{figure}

In Fig. \ref{Fig1_Kink}, we show the raw data in the form of 2D
intensity plots of the spectral function, along with the
corresponding EDC's for ${\vec k}$-space cuts, in the $\Gamma$ to
$(\pi,\pi)$ and the $(\pi,0)$ to $(\pi,\pi)$ directions.  In Fig.
\ref{CloseEDC}, we show the same data represented in a more
traditional fashion with EDC's selected at larger angular
intervals (0.5$^{\circ}$).  In the $\Gamma$ to $(\pi,\pi)$
direction, a large broad feature disperses out of the background,
sharpens to a peak at $E_{F}$, and then disappears.  As pointed
out previously\cite{estruct}, this is behavior not unlike that
found in optimally doped $p$-type compounds.  In the $(\pi,0)$ to
$(\pi,\pi)$ direction, an almost parabolic-shaped band centered
around $(\pi,0)$ is observed. A similar sharpening of the feature
as it disperses to $E_{F}$ is seen here as well, however, the
$|{\vec k}| < |{\vec k_F}|$ EDC's seem to show additional
structure in the form of two separate peaks. This aspect has been
also noted previously\cite{estruct,NCCOGap,Sato}.

\begin{figure}[htb]
    \centering
    \includegraphics[width=7.5cm]{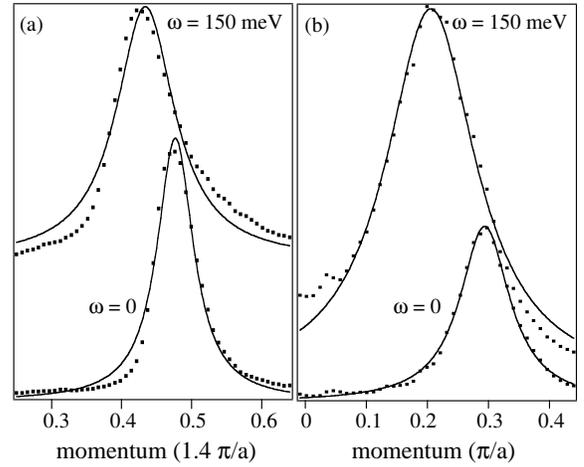}
    \vspace{.1cm}
\caption{MDC's from Fig.  1 at $\omega=0$ and $\omega=150$ meV
binding energy along the two main symmetry directions.  (a)
$\Gamma$$\rightarrow$$(\pi,\pi)$. (b) ($\pi,
0$)$\rightarrow$($\pi,\pi$).  The spectra are offset for clarity.
} \label{Lorentzfits}
\end{figure}

The MDC's from the raw false intensity plots in Fig.
\ref{Fig1_Kink}a and \ref{Fig1_Kink}c were fitted to Lorentzians
as detailed above.  Representative fits are shown in Fig.
\ref{Lorentzfits} at $\omega=0$ and $\omega=150$ meV slices.  At
energies higher than those displayed, the MDC line-shape begins to
differ more significantly from a simple Lorentzian.

\begin{figure}[hbt]
    \centering
    \includegraphics[width=8.5cm]{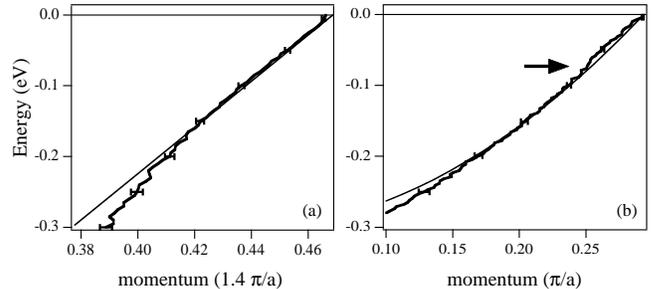}
    \vspace{.1cm}
\caption { Dispersions fitted from the spectral function analysis.
(a) $\Gamma$$\rightarrow$($\pi, \pi$).  The sharp upturn at low
energies ($\omega < 15$ meV) is believed to be an artifact of the
finite energy resolution.  (b) ($\pi, 0$)$\rightarrow$($\pi,
\pi$). The smooth lines are simple fits (linear and parabolic for
(a) and (b) respectively).  The horizontal bars represent the
uncertainty in the peak position from the Lorenztian fits.  The
arrow marks the position of the low-energy kink.}
    \label{dispersion}
\end{figure}

\begin{figure}[h]
    \centering
    \includegraphics[width=8.5cm]{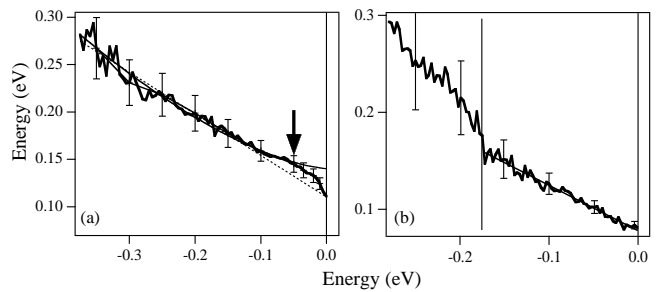}
    \vspace{.1cm}
\caption{Energy widths $\Sigma^{\prime\prime}$ as fitted from
spectral function MDC's and multiplied by velocity.  (a)
$\Gamma$$\rightarrow$($\pi, \pi$) The solid and dashed line are
power -law fits [ to $\Sigma^{\prime\prime}(\omega) =
\Sigma^{\prime\prime}(0) + A\omega^{\alpha}$] with $\alpha=1.55$
and $1$, respectively.  The arrow marks the point of low-energy
drop in $\Sigma^{\prime\prime}$.  (b) ($\pi,
0$)$\rightarrow$($\pi, \pi$). The solid line is a linear fit.  The
vertical bar at 0.18 eV marks the energy beyond which the MDC
width fits are not reliable.  Vertical bars represent the
uncertainty in the fitted Lorenztian width.} \label{Ewidths}
\end{figure}

At the displayed relevant lower energies, the Lorentzian
functional forms fit the MDC's relatively well, showing that,
although there may be large momentum dependence in the self-energy
in other parts of the BZ, the changes across the displayed regions
are minimal at low energy.  The fact that the fits begin to break
down at higher energies indicates that momentum dependent effects
may become relevant.  These fits to the peaks positions are
displayed in Fig. \ref{dispersion} with the corresponding energy
widths $\Sigma^{\prime\prime}$ shown in Fig. \ref{Ewidths}.
Horizontal and vertical errorbars in the Figs. \ref{dispersion}
and \ref{Ewidths}, respectively, represent the uncertainties in
the values from the Lorentzian fits.  In Fig \ref{dispersion}a we
show the dispersion along the zone diagonal plotted from the
momentum space peak position.  As seen by comparison with the
linear fit, there is only a gentle slope change and no sign of a
sharp mass-renormalization at low energy. This is in sharp
contrast to the $p$-type materials where this is the region of the
BZ that shows the most unambiguous effect at $\approx$ 70 meV.

The behavior along the ($\pi, 0$)-($\pi, \pi$) direction differs
from that of the zone diagonal.  Here, the smooth dispersion
deviates at low energy from the parabolic fit in the region
indicated by the arrow in Fig \ref{dispersion}b.  The deviation in
the dispersion is greater than the uncertainty in the fit.  The
small mass renormalization around 55 meV has the same energy scale
as the dip structure in the EDC in Fig. \ref{CloseEDC}. This is
close to the same energy scale as the mass renormalization onset
in the $p$-type compounds.  If one is to identify this apparent
mass change with coupling to a bosonic mode, a simple analysis for
the dimensionless coupling parameter $\lambda ^{\prime} = \frac
{{\vec v_{F}^{0} }} { v_{F} } - 1$ gives a value $\lambda
^{\prime} \approx 0.3$.  This may be compared with values $\approx
0.5$ extracted via the same method from the zone diagonal
direction of the $p$-type materials near optimal doping
\cite{Disclaimer}.  In this simple analysis, we used as a value
for ${\vec v_{F}^{0} }$ the velocity from the dispersion above the
expected kink energy and the ${\vec v_{F} }$ from a low -$\omega$
fit, as was done in the analysis for the $p$-type materials
\cite{Ale}.

Closer to the spirit of how a mass renormalization is typically
defined would be to use a $\omega \rightarrow 0$ extrapolation of
a function fitted to the dispersion at higher energy ($\omega >
100 $ meV).  Analyzing the data in this fashion gives velocities
of ${\vec v_{F}^{0 \medskip eff}(\omega \rightarrow 0) } = 4.3
\times 10 ^{5}$ m/sec (2.3 eV$\cdot a/ \hbar \pi$) for the
$\Gamma$-($\pi, \pi$) FS crossing and ${\vec v_{F}^{0 \medskip
eff}(\omega \rightarrow 0) } = 3.4 \times 10 ^{5}$ m/sec (1.8
eV$\cdot a/ \hbar \pi$) for the ($\pi, 0$)-($\pi, \pi$) FS
crossing.  Using this value for $v_{F}^{0}$ for the ($\pi,
0$)-($\pi, \pi$) cut gives a smaller $\lambda^{\prime}$ of
$\approx 0.1$. However, the values for $\lambda^{\prime}$ near
($\pi, 0$) arrived at in this fashion are not readily comparable
to those of the hole doped compounds along the zone diagonal as
their dispersion there does not appear to recover to something
resembling an unrenormalized dispersion within the experimental
window.  This is most likely a consequence of the fact that for
straightforward electron-boson coupling one expects that the
dispersion recovers to the bare one within about 5 times the
bosonic band width.  In the ($\pi, 0$) region of the BZ, the local
bare electronic band width may be narrower than this bosonic
bound, so one has to regain a parabolic dispersion more quickly.

We note that finite resolution effects should not cause an
inaccurate parametrization except very close to the Fermi energy.
Close to $E_{F}$ where the Fermi function differs significantly
from unity, finite energy resolution ($\Delta E$) weights the
signal at each individual $\omega$ point disproportionately
towards higher energy so that only fits to $\omega$ higher than
$\Delta E$ are reliable and intrinsic.  The sharp change to higher
velocity at $\omega < 15$ meV in the $\Gamma$ to $(\pi,\pi)$
direction is obviously explained by effects of this kind.  As will
be shown later, the data from this energy region show also a sharp
decrease in scattering rate along the zone diagonal.  We consider
this also to be an artifact of finite energy resolution, although
the reason such an effect does not show up in the $(\pi,0)$ to
$(\pi,\pi)$ direction is an open question.

As detailed above, the widths of the MDC Lorentzians can be
parametrized as $\Sigma^{\prime\prime}({\vec k},\omega)/{\vec
v_{F}^{0} }$.  Like the case of $\lambda^{\prime}$, in the absence
of a definitive independent measure for ${\vec v_{F}^{0} }$, we
use for ${\vec v_{F}^{0 \medskip eff} }$ the ${\vec v}$ from a
$\omega \rightarrow 0$ extrapolation of a fit to the higher-energy
dispersion to get a rough estimate for $\Sigma^{\prime\prime}$.
For instance, in the ($\pi, 0$)-($\pi, \pi$) direction, we use the
value for the ${\vec v_{F}^{0} }$ of the fitted parabola.  This
fit, although displayed as extrapolated to $\omega=0$, does not
use input data for $\omega <100$ meV.

Along the $\Gamma$-($\pi, \pi$) direction (Fig.  \ref{Ewidths}a)
$\Sigma^{\prime\prime}(\omega)$ is best fit (solid line) in the
higher energy region (0.4 eV$>\omega>$0.09 eV) to a power law
form, $\Sigma^{\prime\prime}(\omega) = \Sigma^{\prime\prime}(0) +
A\omega^{\alpha}$, with $\Sigma^{\prime\prime}(0) = 0.14$ eV and
$\alpha=1.55$.  Note, that the best fit to a linear dependence
(dashed line) with a $\Sigma^{\prime\prime}(0)$ of 0.11 eV does
not fit nearly as well.  We observe a low-energy drop in
$\Sigma^{\prime\prime}$ at approximately 0.07 eV.  Although the
drop is subtle, we believe the effect to be intrinsic as it was
seen in a number of samples and is outside the uncertainty in the
fits as represented by the error bars.  Note that this is close to
the same energy that the mass renormalization is found at in the
($\pi, 0$)-($\pi, \pi$) direction.  The precipitous drop in
$\Sigma^{\prime\prime}$ for $\omega<15$ meV is found from fitted
MDC's that gave the artificial upturn in the dispersion in Fig.
\ref{dispersion}a and therefore appears to be an artifact of
finite energy resolution.

In contrast, along the ($\pi, 0$)-($\pi, \pi$) direction,
$\Sigma^{\prime\prime}(\omega)$ is best fit by a linear dependence
on $\omega$ with $\Sigma^{\prime\prime}(0) = 0.08$ eV with no
signature of a low-$\omega$ drop in the scattering rate.  Note
that due to the much smaller local electronic band width in this
region of the BZ, the MDC Lorentzian fit breaks down at energies
higher than $\approx$ 0.2 eV as may have been expected from a
visual inspection of Fig.  \ref{Fig1_Kink}c.  It is interesting to
note that $\Sigma^{\prime\prime}(0)$, extracted in this way, is
very different for the two directions.  Also note that the finite
energy ($\Delta E$) and momentum resolution ($\Delta {\vec k }$)
only give a minor broadening contribution to
$\Sigma^{\prime\prime}$ as $\Delta \Sigma^{\prime\prime}= \sqrt{
(\partial E_{{\vec k }}/ \partial {\vec k }  \cdot \Delta {\vec k
} )^{2} + (\Delta E)^{2} } \approx$ 28 meV.

\section{Discussion}

As discussed above and in Ref. \cite{Ale}, due to its ubiquity,
temperature dependence, and energy scale, the strong velocity
renormalizations in the $p$-type materials has been attributed to
an interaction of the charge-carriers with phononic degrees of
freedom.  In the $n$-type compounds we see some evidence of
similar effects although important differences exist. As detailed
above, there is no kink along the zone diagonal, which is the
region of the BZ that shows the most unambiguous effect in the
hole doped materials.  There does appear to be a kink along the
($\pi, 0$) to ($\pi, \pi$) direction.  However, as we will discuss
below, the interpretation of this feature is not straightforward.

If there exists a symmetry reducing order (or associated
fluctuations) with characteristic wave vector ($\pi, \pi$), then
one may expect to see two features in the EDC's in portions of the
BZ where the unreconstructed band structure is below $E_F$ in
momentum regions straddling both sides of the antiferromagnetic BZ
boundary since the effect of such an order is to reflect features
across the antiferromagnetic BZ boundary, i.e., spectral features
from $(h,k)$ would be expected to show up at $(-h + \pi ,-k +
\pi)$. For instance, EDC's from near the Fermi surface near
($\pi,0.3\pi$), would have features reflective of the
unreconstructed bands well below $E_{F}$ from near ($0.7\pi,0$).
In contrast, EDC's from the band near ($0.45\pi,0.45\pi$) will not
have features that derive from near ($0.55\pi,0.55\pi$) as this
region of the BZ in the unreconstructed band structure is well
above $E_{F}$.  In this way, such a BZ reducing order may be
expected to greatly modify the spectral function near ($\pi, 0$),
but not as much near ($\pi/2, \pi/2$) for an unreconstructed
hole-like Fermi surface centered at ($\pi, \pi$).  Two such
features in the EDC can give a kink in the effective dispersion if
an MDC analysis is performed. As it has been previously noted that
there are two features in the EDC's near the ($\pi, 0$)
position\cite{estruct,NCCOGap,Sato} (see Fig. \ref{CloseEDC}b) and
there is evidence for fluctuations of some kind of ($\pi, \pi$)
symmetry reducing order, we do not believe that there is
sufficient evidence to unambiguously assign the ($\pi, 0$)-($\pi,
\pi$) kink feature, found by an MDC analysis, the same origin as
the kink observed in the $p$-type compounds. We believe there is
no means to unambiguously separate the ($\pi, \pi$) symmetry
reducing effect. In this regard, the kink found along ($\pi,
0$)-($\pi, \pi$) in the present work appears to be of a different
sort than the zone diagonal kink found in the $p$-type. The
current case may be an example of a mass renormalization due to
the effects of generic residual antiferromagnetism (i.e. not a
resonance mode) or other symmetry reducing order with
characteristic wavevector ($\pi, \pi$), and therefore its
observation does not directly bear on the debate regarding the
relative merits of the phonon vs magnetic resonance mode scenarios
for the spectral function of the cuprates.

Despite the uncertainties in the MDC analysis one can make a
strong statement about the lack of a kink along the zone diagonal.
By the same condition as given above, in this region of the BZ, we
do not expect to see ($\pi, \pi$) features (nor are any observed)
as the unreconstructed Fermi sea is only filled on one side of
antiferromagnetic BZ boundary. Visual inspection of the MDC fitted
dispersion in the 0.2 eV $>\omega>0$ eV range shows there is no
feature resembling a sharp kink in the dispersion \cite{negative}.

The dependence of the scattering rate on frequency along the zone
diagonal is different from the linear dependence reported in the
optimally doped $p$-type materials\cite{VallaScience} (note that
some groups have found possible deviations from linear
\cite{XJZhou}).  This stronger $\omega^{1.55}$ dependence in the
electron doped compounds compares more closely to the quadratic
temperature dependence of their resistivity that to the linear
temperature dependence of the hole doped materials at optimal
doping \cite{Gollnik,Onose_PG}.

Our results for $\Sigma^{\prime\prime}$ may also be compared to
the linear dependence of the low-energy ($>$1000 cm$^{-1}$)
scattering rate that has been found by an inversion of the optical
conductivity\cite{Homes,Basov_NCCO}.  The conventional wisdom is
that zone diagonal states give the primary contribution to
transport due to their supposedly higher ${\vec v_{F} }$.  At
first glance this is not consistent with our finding here.
However, as we have found for the electron doped material that the
${\vec v_{F}}$ at the $\Gamma$-($\pi, \pi$) FS crossing is not
greatly different from that found at the ($\pi, 0$)-($\pi, \pi$)
FS crossing [this is another consequence of the fact that the
($\pi, 0$) saddle point is $\approx$ 300 meV below $E_{F}$], the
($\pi, 0$) region (where a linear dependence is found) may
contribute somewhat equally to the optical scattering rate.
Moreover, as may be seen in Fig. \ref{Ewidths}a, the scattering
rate along the zone diagonal can be fit approximately by a linear
dependence above 130 meV and it may be that it is this energy
which is contributing the most to the optical excitations for
E$_{\gamma}>$1000 $cm^{-1}$ ($\approx 100$ meV).  Optical
experiments also find a low-energy drop in the scattering rate
below an energy (650 cm $^{-1}$ $\approx$ 70 meV) very close to
that where we find the depression in $\Sigma^{\prime\prime}$ and a
mass renormalization \cite{Homes,Basov_NCCO}.

The energy at which we observe the mass renormalization along the
($\pi, 0$) to ($\pi, \pi$) direction is the same energy, to within
experimental uncertainty, that we observe the drop in the
scattering rate along $\Gamma$ - ($\pi, \pi$).  Although these
energy scales are consistent with each other, this observation is
inconsistent with Kramers-Kronig considerations. Since
$\Sigma^\prime$ and $\Sigma^{\prime\prime}$ are real and imaginary
components derived from the same response function, they should be
related via a Kramers-Kronig transform due to causality.  If
$\Sigma^\prime$ along a particular cut shows a kink, then
$\Sigma^{\prime\prime}$ should show a corresponding drop.  The
reason why such a behavior is not observed is unclear, but may be
related to momentum dependence in the self-energy masking such
effects.  If the self-energy is momentum dependent, the dispersion
and width obtained from an MDC analysis cannot be simply related
to $\Sigma^\prime$ and $\Sigma^{\prime\prime}$, respectively.  It
does seem that the qualitative conclusion that there is little
renormalization along the zone diagonal remains valid. The
situation for the ($\pi, 0$)-($\pi, \pi$) direction is less clear,
as the mass renormalization is present here and the situation to
compare to in the $p$-types near ($\pi, 0$) more uncertain.

As detailed above, it seems reasonable to associate the mass
renormalization on the $p$-type side of the phase diagram with a
strong coupling of charge carriers to phononic degrees of freedom.
Specifically, the principal phonon has been conjectured \cite{Ale}
to be the oxygen half-breathing mode, which is seen to have
anomalous softening and spectral weight changes in the relevant
energy range near the doping induced metal-insulator transition as
probed by neutron scattering
\cite{McQueen1,McQueen2,Braden,Petrov}.  A number of recent
studies have found similar signatures of phononic anomalies in the
electron doped materials \cite{Kang,dAstuto1}. Kang {\it et al.}
found changes with doping in the generalized phonon density of
states around $\approx$ 70 meV by neutron scattering.  d'Astuto
{\it et al.} measured phonon dispersions via inelastic x-ray
scattering and assigned the softening they found in the 55 - 75
meV energy range to the same oxygen half-breathing mode that
anomalies are found in in the $p$-type materials.  These studies
give evidence for phononic effects in the electron doped materials
that are somewhat similar to those found in the hole doped
compounds.

There are a number of differences that do exist in these anomalies
that may provide a starting point for understanding why similar
signatures in the boson spectrum may not cause similar eefects to
appear in ARPES. Although the biggest changes in the phonon
density of states probed by Kang {\it et al.} happen at similar
doping levels in La$_{2-x}$Sr$_{x}$CuO$_4$ and
Nd$_{2-x}$Ce$_{x}$CuO$_4$ ($x \approx 0.04$), the doping levels
are at very different relative positions in the phase diagram,
with $x = 0.04$ being still well into the antiferromagnetic (and
possibly more insulating) phase for the electron doped compound.
As such modifications in the phonon spectrum may be associated
generally with screening changes (and hence electron-phonon
coupling) with the onset of metallicity, this demonstrates the
possibility that the changes in the NCCO phonon spectrum, although
superficially similar in the electron and hole doped materials,
are in some sense different.  d'Astuto {\it et al.} found that
although the softening of the phonon dispersion appears in a
somewhat similar way at a similar energy scale as in the $p$-types
compounds, differences do exist in the shape of the anomaly in the
phonon dispersion. Moreover, on general grounds, since the
purported soft phonon is the oxygen half-breathing mode, one may
naively expect a weaker coupling for this mode with electron
doping, as Madelung potential considerations\cite{Torrance}
indicate that doped electrons will preferentially sit on the Cu
site, whereas doped holes have primarily oxygen character.

Kang {\it et al.} have mentioned \cite{Kang} that it would be
interesting to look for mass renormalizations in the ARPES spectra
of the $n$-type underdoped compounds that they first detect phonon
anomalies in.  However, we believe that if the electron-phonon
coupling is of the same nature in the $n-$ and $p-$ type materials
it should reveal itself in the ARPES spectra of the
highest-$T_{c}$ electron doped samples, as it does in the hole
doped materials.  Moreover, the issue of looking for mass
renormalizations and at peak widths may not even be relevant at
the extremely low dopings ($x=0.04$) where Kang {\it et al.} see
the largest changes in the phonon density of states, as there are
only regions of the BZ with enhanced near-$E_{F}$ spectral weight
and no well-defined electronic peaks\cite{doping}.  The question
of what the electron-phonon coupling parameter is as one enters
the Mott insulating state at very low doping, may not even be
particularly well posed, as the near-$E_{F}$ spectral weight
vanishes as $x=0$ is approached.

\section{Conclusion}

A modern spectral function analysis reveals very different and
likely weaker signatures of possible bosonic effects in the ARPES
spectra of Nd$_{1.85}$Ce$_{0.15}$CuO$_4$ along the two main
symmetry directions of the BZ as compared to the hole doped
compounds. However, unlike the $p$-type compounds, we believe that
one cannot distinguish the various contributions to the
self-energy into separate effects, as the simple analysis yields
results which are not consistent with Kramers-Kronig
considerations.  Despite the complexities inherent in assigning
various features in the spectra to distinct sources, a direct
comparison between the $p$- and $n$-type compounds does have
utility, as we have shown the lack of a mass renormalization along
the zone diagonal.  Our analysis shows that whatever is causing
the distinct effect in the hole doped materials manifests itself
less strongly on the $n$-type side, perhaps due to a weaker
electron-phonon coupling or masking effects from other
interactions.  Although such observations are consistent with a
weaker electron-phonon coupling in the electron doped cuprates, a
strong discrepancy exists between these measurements and a number
of scattering studies that point to strong electron-phonon
coupling.  Perhaps, some of the intrinsic differences between
electron and hole doped materials may provide a route towards
reconciling these different measurements and thereby giving
insight into the underlying phenomenon.

\section{Acknowledgments}

The authors would like to thank J.D. Denlinger for access to his
data analysis routines and M. d'Astuto, A.  Lanzara, and S.A.
Kivelson for helpful conversations.  Experimental data were
recorded at the Advanced Light Source which is supported by the
DOE Office of Basic Energy Science, Division of Materials Science
under Contract No. DE-AC0376SF00098.  Additional support was
through the Stanford Synchrotron Radiation Laboratory which is
operated by the DOE Office of Basic Energy Science, Division of
Chemical Sciences and Material Sciences.  The crystal growth work
at Stanford was supported by the U.S.  Department of Energy under
Contracts No. DE-FG03-99ER45773 and No.  DE-AC03-76SF00515, and by
NSF CAREER Award No.  DMR-9985067.  The crystal growth work in
Tokyo was supported in part by Grant-in-Aids for Scientific
Research from the Ministry of Education, Science, Sports, and
Culture, Japan, and NEDO.


\begin{references}
\bibitem{FermiL}  We should note that this analysis is predicated
on the fact that it is physically reasonable to express
interaction effects as $electronic$ self-energies.  In this sense,
the marginal Fermi liquid phenomenology \cite{Varma} may represent
a ``worst case'' scenario in assigning physical meaning to such
self-energies.  If, in fact, the cuprates are non-Fermi liquids in
the sense that the fundamental excitations have integrity that are
distinct from electrons, this analysis may not be valid.  Indeed,
in such a case or in any case where the MDC's are not well
represented as Lorentzians, this analysis will cause an inaccurate
and unphysical parameterizations of the spectra.
\bibitem{samples} Data consistent with those presented here were
also obtained on samples where the reducing condition was 100 h in
flowing argon gas at 1000 C.
\bibitem{Disclaimer}  These values for $\lambda^{\prime}$ are in
actuality an upper bound on the full renormalization as there is
no independent measure of the bare dispersion to get ${\vec
v_{F}^{0} }$.
\bibitem{negative}  The slight negative curvature in the
dispersion found along the zone diagonal is due to the MDC fits
becoming less reliable at higher energies.  A fit using EDC's at
high energy reveals the expected positive curvature.
\end{references}
\end{document}